\documentclass[12pt]{article}

\usepackage{psfig}
\usepackage{latexsym}
\usepackage{amssymb}
\usepackage{epsfig,amsmath,graphics}

\newfont{\kreuz}{msbm10 scaled\magstep1}
\newfont{\Deutsch}{eufb10 scaled\magstep1}
\newfont{\deutsch}{eufb10}
\newfont{\schreib}{eusm10 scaled\magstep1}

\newcommand{\half}{\frac{1}{2}}
\newcommand{\N}{\mathcal{N}}
\newcommand{\D}{\mathcal{D}}
\newcommand{\Tr}{\mbox{ Tr}}
\newcommand{\tBar}{\bar{\theta}}
\newcommand{\sqt}{\frac{1}{\sqrt{2}}}
\newcommand{\G}{h}
\newcommand{\EtPV}{e^{V}}
\newcommand{\EtMV}{e^{-V}}

\begin{document}
\begin{titlepage}
\begin{flushright}
HUTP-01/A004 \\
LBNL-47410\\
UCB-PTH-01/02\\
hep-th/0101233\\
\end{flushright}
\vskip 2cm
\begin{center}
{\large\bf Higher Dimensional Supersymmetry in 4D Superspace}
\vskip 1cm
{\normalsize
Nima Arkani-Hamed$^{a,b}$, Thomas Gregoire$^b$ and Jay Wacker$^b$\\
\vskip 0.5cm

(a) Jefferson Physical Laboratory\\
Harvard University\\
Cambridge, MA 02138\\
\vskip .3cm

(b) Department of Physics, University of California\\
Berkeley, CA~~94720, USA\\
and \\
Theory Group, Lawrence Berkeley National Laboratory\\
Berkeley, CA~~94720, USA\vskip .1in}
\end{center}
\vskip .5cm

\begin{abstract}
We present an explicit formulation of supersymmetric Yang-Mills theories 
from $\D=$ 5 to 10 dimensions in the familiar $\N=1,\D=4$ superspace. This
provides the rules for globally supersymmetric model building with 
extra dimensions 
and in particular 
allows us to simply write down $\N=1$ SUSY preserving 
interactions between bulk fields and fields 
localized on branes. We present a few applications of the formalism by way of 
illustration, including supersymmetric ``shining'' of bulk fields,
orbifolds and 
localization of chiral fermions, anomaly inflow and 
super-Chern-Simons theories. 

\end{abstract}

\end{titlepage}

\pagebreak

\section{Introduction}

In recent years there has been a resurgence of interest in theories with
extra dimensions, which are in one way or another more accesible than
dimensions compactified at the Planck scale. Working with dimensions larger
than the Planck length allows us to study higher-dimensional physics 
in a ``bottom-up'' approach, within a sensible effective field theory. 
 The new space in extra dimensions has opened up a number of novel approaches to
old questions in beyond the standard model physics. Theories where the SM
fields are stuck to a 3-brane while gravity is free to propagate in extra
dimensions have been used to address the hierarchy problem \cite{ADD,RS}, allowing us to
lower the fundamental scale of gravity, and the ultimate cut-off on
effective field theory, to TeV energies. This also opens up 
the possibility that the SM fields can propagate in extra dimensions of a 
size near the TeV scale \cite{Antoniadis,DDG}. 
Since the higher-dimensional gauge theory becomes
strongly coupled in the UV, it must be embedded in a sensible UV completion
not far above a TeV, and this is possible if the fundamental scale is
itself in this region. 

Many interesting model-building possiblities involve
non-gravitational fields propagating in the extra dimensions. For instance,
sources for massive bulk fields can  ``shine'' an exponentially
falling profile for them in the bulk \cite{shining}, which can be used to explain
small Fermion masses. Another possibility is that different SM 
Fermions can be localized to
different points in the extra dimensions \cite{localized}; their small overlapping
wavefunctions could also lead to a mass hierarchy, or proton
stability. Electroweak symmetry breaking can be triggered by the SM gauge
interactions getting strong in extra dimensions \cite{EWSB}, and  
a number of interesting models for SUSY breaking put the SM fields in the
bulk \cite{gauginomed,martinwitek,Peskin}. 

Many of these mechanisms are generic to the existence of extra dimensions
and have nothing to do with their size per-se: they could in principle 
work just as well with extra dimensions near the GUT scale as near the TeV
scale. However, if these extra dimensions are to be far above the TeV
scale, some physics other than a low fundamental cut-off must be used to
stabilize the electroweak scale. Supersymmetry is a natural
candidate to do this. Then, in order to be able to work with extra
dimensions, we need to know the rules for building supersymmetric theories
in higher dimensions. Furthermore, since many of the models use fields
localized on 3-branes in the extra dimensions,(for instance on D-branes or at orbifold
fixed points), it is also of interest to be able to couple bulk fields to
localized fields in a way preserving at least $\N=1$ SUSY in 4D. 

It is therefore desirable to have systematic rules for writing down supersymmetric
Lagrangians in higher dimensions, allowing supersymmetric couplings to
fields localized on 3-branes. The main work along these lines we are aware of
is the pioneering paper of Peskin and Miraberlli \cite{Peskin}, which showed how
to couple 5D vector and hyper-multiplets to boundaries in a supersymmetric
way, using an off-shell component formalism. However, the formalism is not
familiar to 4D SUSY model-builders, and the extension of the
formalism to higher dimensions is not obvious.

In this paper, we will present a formalism for explicilty constructing
higher-dimensional SUSY theories in a simple way, within the familiar
$\N=1,\D=4$ superspace. The simple observation is that, whatever the
higher-dimensional theories are, they certainly contain the ordinary 4D
SUSY, and therefore they must have an ordinary 4D superspace description. 
The superfield content of the 4D theory is easy to guess, simply by knowing
the total number of SUSY generators in the full theory. For instance in 5D,
the smallest spinor is a Dirac spinor with 8 real components, which means
there are a minimum of 8 supercharges, or $\N=2$ in 4D. From the 4D
viewpoint, we have either hypermultiplets or vector multiplets. Consider
hypermultiplets for simplicity. In $\N=1$ language, they break into two
chiral multiplets $H,H^c$. Furthermore, we have one of these superfields
for each point $x_5$ in the 5'th dimension. So, our field content consist of
superfields $H(x_5),H^c(x_5)$. From the 4D point of view $x_5$ can simply
be thought of as a label. Now, our task is to write down a superspace action for
these fields that, once all auxilliary fields have been integrated out,
reduces to the correct component action for the 5D theory. This is very
easy to do, as the possible terms are heavily constrained by various
symmetries. For this particular example this was done in \cite{SUSYshining}, and will
be reviewed in the next section. We will carry this procedure out for all
globally supersymmetric theories from $\D=5$ to 10 dimensions in this paper. 
But in any case, once we have the action for the bulk theory
written in 4D superspace, it is trivial to couple bulk fields to fields localized
on 3-branes, in a way preserving $\N=1$ SUSY. We simply add 
additional 4D superspace interactions localized at particular
locations in the transverse dimensions. 

We will begin by describing SUSY gauge theories in $5,6$ dimensions, where the field
content is the same as $\N=2$ in 4D. We then move on to the cases $\D=7$ to
10, where the field content is that of $\N=4$ in 4D. For the gauge
multiplets, we 
first discuss the Abelian theory before giving the non-Abelian
generalizations. We then discuss a number of applications in the remainder
of the paper. 

After this work was posted to hep-th, we were informed by A. Sagnotti and
W. Siegel that a superfield formulation of $\D=10$ SYM was given in 
\cite{siegel}. The formulation there is essentially identical to the one we
present for this case. The action given in \cite{siegel} has an extra Wess-Zumino-Witten
type term required to make it fully gauge invariant--this term was missed
in the first version of our paper. However, the new term vanishes in
Wess-Zumino gauge, and so our previous results are unmodified in WZ gauge. 
\cite{siegel} did not discuss the construction of minimally supersymmetric models
in $\D=5,6$, nor the applications of the formalism to brane-bulk couplings
and model-building.

\section{$\D=5,6$}

In $\D=5$ the smallest spinor is a 4 component Dirac spinor
with 8 real degrees of freedom.  In $\D=6$ the smallest 
spinor is a 4 component Weyl spinor with 8 real degrees
of freedom.
Therefore, for $\D=5,  6$ the most simple supersymmetric
theories, those with one copy of the supersymmetry
generators ($\N =1$), will have the same field content
as a $\D=4$ $\N=2$ theory when dimensionally reduced. 

\subsection{Free Hypermultiplets}

The superfield formulation of the $\D=5$ hypermultiplet
has been described in \cite{SUSYshining}. In $\N=1,\D=4$ superspace,
the 5D hypermultiplet consists of a collection of 4D chiral superfields
$H(x_5),H^c(x_5)$ labeled by the 5'th co-ordinate $x_5$.
Its free action is given by
\begin{eqnarray}
\label{eq : 5d_free_hypermultiplet}
S_5^{\mbox{\tiny{Hyp.}}} &=& \int\!d^5x\Big\{
\int\!d^4\theta \left( \bar{H}^c H^c + \bar{H} H \right) + 
\left(\int\!d^2\theta H^c\left( \partial_5 + m \right) H +\mbox{h.c.}\right)
\Big\}
\end{eqnarray}
Expanding in components and integrating out the auxilliary $F$ components,
the action
(\ref{eq : 5d_free_hypermultiplet}) describes an $\N =1$ $\D=5$ supersymmetric theory containing two complex scalar and one Dirac fermion $\Psi_5^{T} = (\psi,\bar{\psi^c})$
composed of
the 2 component fermions $\psi$ and $\bar{\psi^c}$:
\begin{multline}
S_5^{\mbox{\tiny{Hyp.}}} = -\partial_M H^{\dagger}\partial^M H   -\partial_M {H^c}^{\dagger} \partial^M H^c
 -i \bar{\psi} \bar{\sigma}^{m} \partial_{m} \psi -i \bar{\psi^c} \bar{\sigma}^{m} \partial_{m} \psi^c  - \psi^c \partial_5 \psi - \bar{\psi^c} \partial_5 \bar{\psi}\\
-  m^2 (|{H^c}|^2 + |H|^2) - m (\psi^c \psi + \bar{\psi^c} \bar{\psi}) \\
=
 -\partial_M H^{\dagger}\partial^M H  -\partial_M {H^c}^{\dagger} \partial^M H^c - m^2 (|{H^c}|^2 + |H|^2) + \bar{\Psi}_5(i \gamma^M \partial_M - m) \Psi_5   \\
\end{multline}
Here and throughout the paper, the capitalized indices run over $0,1,2,3,5$ while the lower-case ones run over $0,1,2,3$

In $\D=6$ there are only Weyl and Dirac fermions, so the
smallest multiplet contains a left or right
Weyl fermion.  The action for massless hypermultiplets is.

\begin{eqnarray}
\nonumber
S_6^{\mbox{\tiny{L. Hyp.}}} &=&\int\!d^6x\Big\{
\int\!d^4\theta \left(
\bar{H}^c_L H^c_L +\bar{H}_L H_L \right)
+\left( \int\!d^2\theta  H^c_L\; \partial  H_L +\mbox{h.c.}\right)
\Big\}\\
S_6^{\mbox{\tiny{R. Hyp.}}} &=&\int\!d^6x\Big\{
\int\!d^4\theta \left(
\bar{H}^c_R H^c_R +\bar{H}_R H_R \right)
+\left(\int\!d^2\theta  H^c_R\; \bar{\partial}  H_R +\mbox{h.c.}\right)
\Big\}
\end{eqnarray}
with
\begin{eqnarray}
\begin{array}{ll}
z= \frac{1}{2}(x_5 + ix_6) & \bar{z} = \frac{1}{2} (x_5 -i x_6)\\
\\
\partial = \frac{\partial}{\partial z} = \partial_5 - i \partial_6&
\bar{\partial} = \frac{\partial}{\partial \bar{z}} = \partial_5 + i \partial_6
\end{array}
\end{eqnarray}
Note that the rotational invariance of the transverse 2 dimensional space is realized as
$z \to e^{i \theta} z,H_L^{(c)} \to e^{i \theta/2} H_L^{(c)},H_R^{(c)} \to e^{-i \theta/2} H_R^{(c)}$.
Therefore, unlike the 5D case, we can not make a massive hypermultiplet out of just e.g. $H_L,H_L^c$.
Instead we must combine
a copy of each of the massless hypermultiplets.
\begin{eqnarray}
\nonumber
S_6^{\mbox{\tiny{Hyp. Massive}}} &=& \int\!d^6x\Big\{
\int\!d^4\theta \left(
\bar{H}^c_L H^c_L +\bar{H}_L H_L + \bar{H}^c_R H^c_R + \bar{H}_R H_R \right)\\
\label{eq : 6D_massive}
&&+\left( \int\!d^2\theta
\left( \begin{array}{cc} H^c_R & H^c_L \end{array} \right)
\left( \begin{array}{cc}
m &  \bar{\partial} \\  \partial & m
\end{array} \right) 
\left( \begin{array}{c} H_L\\  H_R \end{array} \right) 
+ \mbox{h.c.} \right)
\Big\}
\end{eqnarray}
The Dirac spinor is now
\begin{eqnarray}
\Psi^{\mbox{\tiny{Dirac}}}_6 = \left(\begin{array}{c} 
\Psi^{\mbox{\tiny{Left Weyl}}}_6\\
\Psi^{\mbox{\tiny{RightWeyl}}}_6
\end{array}\right). 
\end{eqnarray}
Note that the superpotential term is just the Dirac operator in the
transverse 2D space. 

\subsection{Abelian Gauge Theory}

The first step in formulating the higher dimensional 
theories in terms of ordinary $\D=4$, $\N=1$ superspace
is to identify the correct superfields for the
theory.  The $\D=5$ super Yang-Mills theory will have
a 5-vector gauge field, a 4 component Dirac gaugino, and
a scalar.    When dimensionally reduced down to $\D=4$, the
gauge field becomes a 4-vector and a scalar, the gaugino
splits into two Majorana gauginos, and the scalar is
unaffected. So we must have a vector multiplet and chiral multiplet.
This is also obvious since there are 8 real supercharges in
5D, which translates to $\N = 2$ SUSY in 4D, with the $\N = 2$ vector multiplet
composed on an $\N=1$ vector and chiral multiplet.

The correct identification of the fields inside the vector field $V(x_5)$ and chiral
field $\phi(x_5)$  is (with $V$ in the Wess-Zumino gauge, and
$\phi$ in the $y$-basis) 

\begin{eqnarray} 
V &=& -\theta \sigma^m \tBar A_{m} 
+ i \tBar^2 \theta \lambda_1
-  i \theta^2 \tBar \bar{\lambda}_1
+\half\tBar^2 \theta^2 D\\
\nonumber 
\phi &=& \frac{1}{\sqrt{2}}\left(\Sigma + i A_5\right) 
+ \sqrt{2}\theta \lambda_2 + \theta^2 F
\end{eqnarray}
In the above and for the rest of the paper, the dependence of the 4D superfields on the
extra co-ordinates is implicit.
We also demand full $\D=5$ gauge invariance of the
theory.  The gauge transformations of these 2 superfields are:
\begin{align}
V &\rightarrow V + \Lambda +\bar{\Lambda}\\
\nonumber \phi &\rightarrow \phi + \sqrt{2} \partial_5 \Lambda 
\end{align}
The subset of these transformations that preserve the Wess-Zumino gauge, 
correspond exactly to the ordinary $\D=5$ gauge transformations.

It is easy to find a gauge invariant action
\begin{equation}
S_5^{\mbox{\tiny{A}}}= 
 \int\!d^5x\left[ \frac{1}{4 g^2} \int d^2 \theta \;
W^{\alpha}W_{\alpha} + \mbox{h.c.}
+ \int d^4 \theta \frac{1}{g^2}(\partial_5 V - \frac{1}{\sqrt{2}}(\phi + \bar{\phi}))^2\right]
\end{equation}
The first term is familiar and obviously gauge invariant; the second
term is also clearly invariant under gauge tranformations with the variation
of $\partial_5 V$ being canceled by that of $\phi + \bar{\phi}$.

While the $\N =1$ SUSY is manifest in this Lagrangian, the full SUSY and 
higher-dimensional Lorentz invariance is not. To see this, we  expand the 
Lagrangian in components in Wess-Zumino gauge.
Keeping only the Bosonic fields the Lagrangian becomes
\begin{eqnarray}
\nonumber
&& -\frac{1}{4 g^2} F_{\mu \nu}F^{\mu \nu} + \frac{1}{2g^2} D^2 \\
\nonumber
&& -\frac{1}{2g^2} \partial_5 A_\mu \partial_5 A^\mu + \frac{1}{g^2} \partial_5 A_\mu \partial^\mu A_5 - \frac{1}{g^2}\Sigma \partial_5 D \\
&& -\frac{1}{2g^2} (\partial_\mu \Sigma \partial^\mu \Sigma + \partial_\mu A_5 \partial^\mu A_5).
\end{eqnarray}
We can integrate out the auxiliary field $D$ by setting it to its equation of motion, which is
\begin{equation}
D = - \partial_5 \Sigma
\end{equation}
The Lagrangian then naturally arranges itself into the form
\begin{equation}
-\frac{1}{4g^2} F_{MN} F^{MN} - \frac{1}{2g^2} \partial_M \Sigma \partial^M \Sigma
\end{equation}
which is precisely the bosonic part of the 5D vector multiplet composed of the vector field
$A_M$ and the real scalar $\Sigma$. The full Lagrangian including the fermions clearly also works out
correctly.

This superspace form of the 5D Lagrangian is very simple but does not straightforwardly generalize to higher dimensions,
because $\phi,\bar{\phi}$ will transform oppositely
under rotations in the transverse space. We can however re-write the action as
\begin{equation}
\label{5Dgauge}
S_5^{\mbox{\tiny{A}}}=\int d^5 x \left[ \frac{1}{4 g^2} \int d^2 \theta \;
W^{\alpha}W_{\alpha} + \mbox{h.c.} +
\int d^4 \theta \frac{1}{g^2}
\left( (\sqrt{2} \partial_5 V - \bar{\phi})
(\sqrt{2} \partial_5 V - \phi) - \partial_5 V \partial_5 V \right)
\right]
\end{equation}
The gauge invariance of the second and third terms are not as manifest in this form, but it is easy to check.
Under a gauge transformation,
\begin{eqnarray}
\nonumber
\int d^4 \theta (\sqrt{2} \partial_5 V - \bar{\phi})
(\sqrt{2} \partial_5 V - \phi) &\to& \int d^4 \theta ( \sqrt{2} \partial_5 V - \bar{\phi} +
\sqrt{2} \partial_5 \Lambda)(\sqrt{2} \partial_5 V - \phi + \sqrt{2} \partial_5 \bar{\Lambda})\\
= \int d^4 \theta (\sqrt{2} \partial_5 V - \bar{\phi})
(\sqrt{2} \partial_5 V - \phi) &+&
\int d^4 \theta \left[2 \partial_5 V \partial_5(\Lambda + \bar{\Lambda}) + 2 \partial_5 \Lambda \partial_5 \bar{\Lambda}\right]
\end{eqnarray}
where we have used the fact that purely chiral or anti-chiral terms vanish under the full superspace integration.
Similarly,
\begin{eqnarray}
\nonumber
&&-\int d^4 \theta (\partial_5 V)^2 \to -\int d^4 \theta (\partial_5 V + \partial_5 \Lambda + \partial_5 \bar{\Lambda})^2 \\
= && -\int d^4 \theta (\partial_5 V)^2 - \int d^4 \theta \left[2 \partial_5 V \partial_5 (\Lambda + \bar{\Lambda}) - 2 \partial_5 \Lambda \partial_5 \bar{\Lambda} \right]
\end{eqnarray}
so that the sum of the last two terms in eqn. (\ref{5Dgauge}) is gauge invariant.

The extension to $\D=6$ is simple. The transverse
rotational invariance is useful as a guide to constructing the action.
$z$ transforms as $ z \to e^{i \theta} z$, and we suspect that
$\phi$ will combine with $\partial$
to form a covariant derivative so we 
define $\phi$ to transform as $\phi \to e^{- i \theta} \phi$. $V$ is neutral.
The gauge transformations are
\begin{align}
V &\rightarrow V + \Lambda +\bar{\Lambda} \\
\nonumber \phi &\rightarrow \phi + \sqrt{2} \partial \Lambda \\
\end{align}
The 6D action is then the obvious extension of the 5D one:
\begin{equation}
S_6^{\mbox{\tiny{A}}}=\int d^6x \left[
+\frac{1}{4 g^2}\int d^2\theta\;W^{\alpha}W_{\alpha} + \mbox{h.c.}
+ \int d^4 \theta \frac{1}{g^2} \left(( \sqrt{2} \bar{\partial} V -\bar{\phi})
( \sqrt{2} \partial V - \phi) - \partial V \bar{\partial} V \right)
\right]
\end{equation}
In this case the lowest component of the superfield $\phi$ is

\begin{eqnarray}
\phi\,|_{\theta=\bar{\theta}=0} = \sqt A= \sqt\left(A_6 + i A_5\right). 
\end{eqnarray}
This expression reproduces the $\D=5$ super Yang-Mills action when
all dependence on $x_6$ is eliminated and identifying $A_6$ as
the scalar, $\Sigma$, of the $\D=5$ super Yang-Mills theory. 
The auxiliary field $D$ is now proportional to $F_{56}$:
\begin{eqnarray}
D = - \frac{1}{2} \left(\partial \bar{A} + \bar{\partial} A \right) 
= - \left( \partial_5 A_6 - \partial_6 A_5 \right)= -F_{56}
\end{eqnarray}

\subsection{Non-Abelian Theory}
We now generalize to the case of a non-Abelian theory.
Since $\phi$ contains the components of the higher dimensional
gauge field, it must transform in the adjoint.
With the definitions:
\begin{eqnarray}
\begin{array}{cc}
\G = e^{-\Lambda}& 
\bar{\G}= e^{-\bar{\Lambda}}
\end{array}
\end{eqnarray}
the gauge transforms become
\begin{eqnarray}
\begin{array}{ll}
\phi \rightarrow \G^{-1} (\phi - \sqrt{2} \partial) \G&
\EtPV \rightarrow \G^{-1} \EtPV \bar{\G}^{-1}
\end{array}
\end{eqnarray}
where $\phi \equiv \phi^a T^a$ and $V \equiv V^a T^a$. 

The natural guess for the non-Abelian action would be to simply insert various factors of $e^V$: 
\begin{eqnarray}
\nonumber
&&S^{\mbox{\tiny{NA}}}_6= \int d^6x \Big \{ 
\frac{1}{4 k g^2} \Tr \left[ \int\!d^2 \theta\; W^{\alpha}W_{\alpha} 
+ \mbox{h.c.} \right] \\ 
\label{eq : non-Abelian_6D}
&&+ \int\!d^4 \theta 
\; \frac{1}{k g^2} \Tr \left[
(\sqrt{2} \bar{\partial} + \bar{\phi})\EtMV
(- \sqrt{2} \partial + \phi) \EtPV
+ \bar{\partial} \EtMV \partial \EtPV \right] \Big \}
\end{eqnarray}
where Tr$T^a T^b = k \delta^{ab}$.  
This action reproduces the $\D=6$ non-Abelian  super Yang-Mills theory in 
Wess-Zumino gauge. However, this action is not fully gauge invariant under
the gauge transformation. As pointed out in \cite{siegel}, 
we need to add one more term to make it perfectly gauge invariant:
\begin{equation}
\int d^6 x \int d^4\theta \frac{1}{k g^2} \Tr \left[ \bar{\partial}V \frac{\sinh L_V -L_V}{L_V^2} \partial V
\right]
\end{equation}
This has the structure of a WZW term. 
We refer to \cite{siegel} for details on the variation of this term. 
Here we note that the term is absent in $\mathcal{D}=5$ and in all cases, 
vanishes in Wess-Zumino gauge. 
Therefore, one can use (\ref{eq : non-Abelian_6D}), together with any
desired couplings to brane fields, and obtain the correct Lagrangian in
Wess-Zumino gauge. 
To find the $\D=5$ non-Abelian theory, 
one removes the $x_6$ dependence.  This action also reproduces
the appropriate Abelian theories when all fields commute. 

The auxiliary field, $D$, in $\D=6$ again becomes the
higher dimensional field strength, only this time it
is the non-Abelian field strength $F_{56}$:
\begin{eqnarray}
D = - \frac{1}{2}\left( \bar{\partial} A + \partial \bar{A}
+ [ A, \bar{A}]\right) = -F_{56}.
\end{eqnarray}
We can dimensionally reduce this to $\D=5$ to get
\begin{eqnarray}
D = -  \left( \partial_5 \Sigma + i [ \Sigma, A_5] \right)
= - D_5 \Sigma ,
\end{eqnarray}
with $D_5$ being $x_5$ component of the covariant derivative.

\subsection{Coupling to Hypermultiplets}

It is easy to extend our action for free hypermultiplets to the case where
they are charged under a gauge symmetry. With the hypermultiplets belonging to a
representation $R$ of the gauge group $G$, we have the gauge transformations:
\begin{eqnarray}
\nonumber H &\rightarrow& \G H\\
H^c &\rightarrow& \G^c{}^{-1} H^c ,
\end{eqnarray}
with $\G=e^{-\Lambda^a T^a_R}$ and $\G^c = (\G^{-1})^T = (e^{\Lambda^a T^a_R})^T$.
The generalization of our previous hypermultiplet action is trivial; we
simply replace the ordinary $\partial_5$ derivatives with the covariant derivative 
$\partial_5 - \frac{1}{\sqrt{2}}  \phi$:
\begin{eqnarray}
\nonumber
&S_5^{\mbox{\tiny{Hyp. Gauge}}}
=\int d^5 x \Big\{
\int d^4\theta [H^c \EtPV \bar{H}^c + \bar{H}\EtMV H]\\
& +\left[\int d^2 \theta (H^c (m + (\partial_5 - \frac{1}{\sqrt{2}}\phi)) H) 
+ \mbox{ h.c.} \right]
\Big\}
\end{eqnarray}
(of course here $V = V^a T^a_R, \phi = \phi^a T^a_R$).    

In $\D=6$ we must choose our gaugino to be either a left or right 
handed Weyl field. To make a covariant derivative
we must combine the left-handed $\phi$ with $\partial$ and the 
right-handed field with $\bar{\partial}$. 
Therefore, hypermultiplets of
a given handedness can not couple to gauge fields of opposite
handedness.  The action for a $\D=6$ hypermultiplet coupled to a gauge field
of the same handedness is

\begin{eqnarray}
\nonumber
&&S_6^{\mbox{\tiny{Hyp. Gauge}}}
=\int d^6 x \Big\{
\int d^4\theta [H^c \EtPV \bar{H}^c + \bar{H}\EtMV H]\\
&& +\left[\int d^2 \theta\; \sqrt{2}H^c (\partial - \frac{1}{\sqrt{2}}\phi) H
+ \mbox{ h.c.} \right]
\Big\}
\end{eqnarray}

\section{$\D=7$ to $10$}

Spinors in $\D=7$ to $10$ dimensions have a minimum of 16 real components.
This means that there are a minimum of  16 real supercharges and thus all theories
in these dimensions must be constructed out of $\N=4$ multiplet
in the 4D language. 
The $\N = 4$ vector multiplet decomposes under $\N = 1$ as 3 chiral
multiplets $\phi_i$ and a vector multiplet $V$, so we need to build our
superspace Lagrangian out of these fields. 

We will only consider the $\D=10$ theory because it is easy to
dimensionally reduce to $\D= 7 \mbox{ to } 9$.
It will be convenient to use complex coordinates,$z^i$, for the transverse space
with 
\begin{eqnarray}
\begin{array}{lll}
z^1 =\frac{1}{2}(x_5 + i x_6)&
z^2 =\frac{1}{2} (x_7 +i x_8)&
z^3 =\frac{1}{2}(x_9 + i x_{10})
\end{array}.
\end{eqnarray}
The transverse rotational invariance is the $SO(6)$ rotating the $x_5,
\cdots, x_{10}$ into each other. The $SU(3)$ subgroup rotating the $z_i$ 
will be useful in constructing invariant actions. 
We will use the convention that $\bar{z}_i = (z^i)^\dagger$, and that 
$\bar{\phi}^i = (\phi_i)^\dagger$.  

Again we will find that the higher dimensional components
of the gauge field will be the lowest components of
$\phi_i$:

\begin{eqnarray}
\phi_j\;|_{\theta=\bar{\theta}=0} &=& \sqt A_j = \sqt ( A_{4+2j} + i A_{3+2j})\\
\nonumber
j &\in& \{1,2,3\}
\end{eqnarray} 
This choice of the embedding was to make the dimensional
reduction from $\D=10$ most transparent. 

\subsection{Abelian Theory}

The appropriate gauge transformation for this theory are
\begin{eqnarray}
\nonumber V &\rightarrow& V + ( \Lambda + \bar{\Lambda})\\
\nonumber \phi_i &\rightarrow&  \phi_i  + \sqrt{2} \partial_i\Lambda.
\end{eqnarray}

The K\"ahler potential of the theory is the natural generalization
of the $\D=5,6$ theory.  However, this will not reproduce the correct
theory.  We need to introduce a superpotential that will complete
the gauge potential kinetic term. This is also obvious since, if we reduce
the theory to 6D eliminating the dependence on $x_7,\cdots,x_{10}$,
$\phi_2,\phi_3$ form a hypermultiplet in 6D, and as we have seen 
the hyper-multiplet kinetic term is completed by a
superpotential term. In any case, 
the $SU(3)$ symmetry and the known result when reduced to $D=6$ specifies 
everything, and we have for the action

\begin{eqnarray}
S_{10}^{\mbox{\tiny{A}}} &=& \int d^{10} x \Big\{\int d^2 \theta \left( \frac{1}{4g^2}W^\alpha W_\alpha
+ \frac{1}{2 g^2} \epsilon^{ijk} 
\phi_i \partial_j \phi_k
\right) + \mbox{h.c.} \nonumber \\
&+&\int d^4 \theta \frac{1}{g^2} \left[\left( \sqrt{2} \partial_i V -  \phi_i \right)
 \left( \sqrt{2} \bar{\partial}^i V - \bar{\phi}^i \right) 
 - \partial_i V \bar{\partial}^i V \right] \Big\}
\end{eqnarray} 
Note that the gauge variation of the superpotential vanishes via
integration by parts and the antisymmetry of $\epsilon^{i j k}$. 
The auxiliary fields $F_i,D$ are given by

\begin{eqnarray}
\nonumber
D &=& - \frac{1}{2}( \partial_i \bar{A^i} + \bar{\partial}^i A_i )\\
{F^i}^{\dagger} &=& -\frac{1}{\sqrt{2}} \epsilon^{ijk} \partial_j A_k .
\end{eqnarray}

\subsection{Non-Abelian Theory}

The non-Abelian action is the natural generalization of the
Abelian one.  The superpotential must be modified
to make it gauge invariant, which is accomplished by replacing the
$\partial_j \phi_k$ with $\partial_j \phi_k -
[\phi_j,\phi_k]/3\sqrt{2}$. 
The gauge transformations are 
\begin{eqnarray}
\begin{array}{ll}
\phi_i \rightarrow \G^{-1} (\phi_i - \sqrt{2} \partial_i) \G&
\EtPV \rightarrow \G^{-1} \EtPV \bar{\G}^{-1}
\end{array}
\end{eqnarray}

\begin{eqnarray}
S^{\mbox{\tiny{NA}}}_{10} &=& \int d^{10}x \Big\{
\int d^2 \theta 
 \Tr\left( \frac{1}{4 k g^2}W^\alpha W_\alpha
+ \frac{1}{2 k g^2} \epsilon^{ijk} \phi_i(\partial_j \phi_k -
 \frac{1}{3\sqrt{2}}[\phi_j,\phi_k]) \right) \nonumber \\ 
& + & 
\int d^4 \theta\; \frac{1}{k g^2} \Tr \left(( \sqrt{2} \bar{\partial}^i + \bar{\phi}^i) \EtMV
( -\sqrt{2} \partial_i + \phi_i)  \EtPV
+  \bar{\partial}^i  \EtMV \partial_i \EtPV  \right) \Big\} \\
&+&
\nonumber
\mbox{WZW term}
%\int d^4 \theta \frac{1}{k g^2} \Tr\left(\bar{\partial}^i V\frac{\sinh L_V -L_V}{L_V^2} \partial_i V\right)
\end{eqnarray}
Once again, the last term vanishes in W-Z gauge. Note that the 
superpotential has the structure of a 
Chern-Simons
term. Under a gauge transformation, the superpotential transforms as 
\begin{eqnarray}
\mbox{Tr} \epsilon^{ijk} \phi_i(\partial_j \phi_k +
 \frac{1}{\sqrt{2}}[\phi_j,\phi_k])
&\to& 
\mbox{Tr} \epsilon^{ijk} \phi_i(\partial_j \phi_k +
 \frac{1}{\sqrt{2}}[\phi_j,\phi_k]) \nonumber \\ &-& 2 \sqrt{2} \mbox{Tr} \left[\epsilon^{i j k} 
(\partial_i \G) 
\G^{-1}(\partial_j \G) \G^{-1}(\partial_k \G) \G^{-1}\right]
\end{eqnarray}
The last term is a total derivative and is the Pontryagin density.
For the transformations that preserve WZ gauge,  
$\Lambda = \exp( i \theta \sigma^m \bar{\theta} \partial_m) a(x)$ 
with no higher components, the Pontryagin term
vanishes identically under the superspace integration.

Finally some brief comments on the $R$ symmetry of these theories. 
In $\D=10$, the  transverse rotational symmetry is $SO(6)$ which is 
homomorphic to $SU(4)$. 
This SU(4) symmetry is the R symmetry of the $\D=4$ $\N=4$ theory.
The superpotential and K\"ahler terms we have written have only an 
explicit SU(3) symmetry.
This SU(3) is a subgroup of the $SU(4)_R$, keeping $\N=1$ SUSY manifest. 
By writing our theory in terms of $\N=1$ superfields,
we choose a special supersymmetry generator and break
the manifest $SU(4)$ R symmetry.
We maintain an $SU(3)$ subgroup
which transforms the three supersymmetry generators
that are orthogonal to our $\N=1$ SUSY generator. 

\section{Some applications}

\subsection{Coupling to sources}

In \cite{shining}, the ``shining'' of bulk massive fields by sources localized
on branes was considered as a mechanism for producing small parameters 
on the brane. This is a consequence of the exponentially small profile for 
the massive field in the bulk, which is given by the massive Yukawa 
propagator. It is natural to try and extend this mechanism to
supersymmetric
theories. This was done in \cite{SUSYshining} for the case of 5D theories, as we 
review below. A source was added to a massive bulk hypermultiplet of the 
form

\begin{eqnarray}
\int d^2 \theta dx_5 \delta(x_5) J H^c
\end{eqnarray}
The F-flatness conditions become:
\begin{eqnarray}
 -F^{\dagger} &=& (m- \partial_5) H^c = 0 \\
\nonumber
 -{F^c}^{\dagger} &=& J\delta(x_5) + (m + \partial_5) H =0 
\end{eqnarray}
These equations have solution $H^c = 0$ and:
\begin{eqnarray}
H = -\theta(x_5)J e^{-m y}
\end{eqnarray}
in infinite space and
\begin{eqnarray}
H = \frac{ - J e^{-m y}}{ 1 - e^{-2 \pi m R}}
\end{eqnarray}
on a circle of radius $R$.

We can do a similar thing for the free 
hypermultiplets in 6D . We add:
\begin{eqnarray}
\int d^6x d^2 \theta \delta(x_5)\delta(x_6) J H_L^c 
\end{eqnarray}
The F flatness conditions are:
\begin{eqnarray}
F_L^{\dagger} &=& \partial H_L^c + m H_R^c  = 0 \\
\label{eq : Fflatness_massive_hyper_lc}
{F_L^c}^{\dagger} &=& -\partial H_L+ m H_R - J \delta(z \bar{z}) = 0 \\
\label{eq : Fflatness_massive_hyper_rc}
{F_R^c}^{\dagger} &=& -\bar{\partial} H_R +m H_L = 0 \\
{F_R}^{\dagger} &=& \bar{\partial} H_R^c + m H_L^c = 0 
\end{eqnarray}
Consider first the massless case $m=0$ and $H_R^c = H_L^c = H_R = 0$; then we have 
\begin{eqnarray}
\nonumber
\label{eq : profile_6D_massless_eqn}
\partial H_L = - J \delta(z \bar{z})
\end{eqnarray}
which has solution:
\begin{eqnarray}
\label{eq : profile_6D_massless}
H_L = - J\frac{\theta(z \bar{z})}{\bar{z}} = -J\frac{\theta(x_5^2 + x_6^2)}{x_5 - i x_6} = -J \frac{e^{i\phi}}{r} \theta(r^2)
\end{eqnarray} 
in infinite space. In order to find the solution on a compact space, say
a torus, we could use the method of images. 

In the massive case, we take $H_{L,R}^c = 0$ and combine equations (\ref{eq : Fflatness_massive_hyper_lc}) and (\ref{eq : Fflatness_massive_hyper_rc}) to get:
\begin{eqnarray}
\bar{\partial} \partial H_R - m^2 H_R &=& -m J \delta^2(z\bar{z}) \\
\nonumber
H_L &=& \frac{1}{m} \bar{\partial} H_R 
\end{eqnarray}
The first equation, is just the Klein-Gordon equation in 2D, so the
solution 
is the Yukawa potential in 2D. For large
$mr$, we have  
\begin{eqnarray}
H_R \sim -J m  e^{-mr}, H_L = -Jm e^{-mr} e^{i \phi}
\end{eqnarray}

It is interesting that $H_L$ acquires a ``vortex'' profile in the 
transverse two dimensions. Even if all the parameters in the Lagrangian 
are real, this vortex profile breaks CP. If the 
Standard Model Yukawa couplings arise through shining via branes that do
not all fall on a straight line, the phase in $H_L$ can be used to
introduce CP violation into the SM in an amusing way.

\subsection{Charged matter on Branes}

Using our formalism, it is very easy to couple bulk gauge fields to charged 
matter on boundaries. We simply add the following term to the appropriate 
higher-dimensional action:

\begin{eqnarray}
\int d^4x d^4 \theta \bar{X} e^{-V}|_{z=0} X 
\end{eqnarray}
where $X$ is a 4 dimensional chiral superfield living on a brane and $z$ represents the extra dimensions. For example, in 5D Abelian case, the action would be:
\begin{multline}
\text{5D free action} + \text{4D free action}\\
 + \frac{1}{g^2}\int d^4 x \Bigg[ A^n\left[-\frac{1}{2} \bar{\lambda_X} \bar{\sigma}^n \lambda_X -\frac{i}{2}\bar{A_X} \partial_n A_X\right]
+ i \sqt  \left(A_X \bar{\lambda_X} \bar{\lambda_1} + h.c\right) \\
- \frac{1}{4}A_n A^n \bar{A_X} A_X 
- \frac{1}{2} D \bar{A_X} A_X \Bigg]
\end{multline}
where $A_X$ is the scalar component of the $X$ multiplet. These are just the usual couplings of a 4D chiral superfield with a 4D vector superfield. But, in our case, the $D$ term is different. Let's examine the D part of the Lagrangian in detail: 
\begin{eqnarray}
\nonumber
\mathcal{L}_D = \frac{1}{g^2}\left(\half D^2 +  D \partial_5 \Sigma - \frac{1}{2}\bar{A}_X A_X D \delta(x_5)\right)
\end{eqnarray}
The first two terms come from the free 5D action part.
Upon eliminating $D$ we get
\begin{eqnarray}
\nonumber
\mathcal{L}_D &=& -\frac{1}{2g^2}\left(\partial_5 \Sigma -\frac{1}{2} \bar{A_X} A_X  \delta(x_5)\right)^2\\
&=& -\frac{1}{2g^2}\left((\partial_5 \Sigma)^2 - \partial_5 \Sigma \bar{A_X} A_X \delta(x_5) + \frac{1}{4}(\bar{A_X} A_X)^2 \delta(0) \delta(x_5)\right) .
\end{eqnarray}
This result was obtained earlier in \cite{Peskin}, but our derivation makes
the ease of the superspace formalism transparent. It is also trivial to 
extend the result to higher dimensions, for instance in 6D we have

\begin{eqnarray}
\mathcal{L}_D &=& -\frac{1}{2g^2}\left(F_{56} - \frac{1}{2} \bar{A_X} A_X  \delta(z)\right)^2\\
\nonumber
&=& -\frac{1}{2g^2}\left(F_{56}^2 - F_{56} \bar{A}_X A_X \delta^2(z) +\frac{1}{4} (\bar{A_X} A_X)^2 \delta^2(0) \delta^2(z)\right)
\end{eqnarray}
where $F_{56} = (\partial_5 A_6 - \partial_6 A_5)$. We note that if
$F_{56}(z=0) \ne 0$ then we get SUSY breaking soft scalar masses 
proportional to the strength of the magnetic field 
on the brane.

\subsection{Orbifolds}
Our formalism is also useful for constructing field-theoretic orbifolds \cite{Orbifold1} 
preserving $\N=1$ SUSY in 4D. Such constructions are useful both for 
obtaining chiral fermions as well as reduced supersymmetry in the
low-energy
4D theory. 

The simplest canonical example is the 
$S_1/Z_2$ orbifold in the 5D case \cite{Orbifold2}.  Consider as an example a $U(1)$ gauge
theory in 5D. It is trivial to see that our 5D action 
is invariant under 

\begin{eqnarray}
V(x_{\mu},x_5) &\rightarrow& V(x_{\mu},-x_5) \\
\nonumber
\phi(x_{\mu},x_5) &\rightarrow& -\phi(x_{\mu},-x_5)
\end{eqnarray}
To construct the orbifolded model, we only keep states that are invariant
under the symmetry, as well as periodic under $x_5 \to x_5 + 2 L$.
That is we impose 
\begin{eqnarray}
V(x_{\mu},x_5) &=& V(x_{\mu},-x_5) \\
\nonumber
\phi(x_{\mu},x_5) &=& -\phi(x_{\mu},-x_5)
\end{eqnarray}
as well as  
\begin{eqnarray}
V(x_{\mu},x_5+ 2L) &=& V(x_\mu,x_5) \\
\nonumber
\phi(x_{\mu},x_5 + 2L) &=& \phi(x_\mu,x_5)
\end{eqnarray}
The physical space is then the interval $[0,L]$. In order to obtain the
low-energy theory, we only need to look at $x_5$ independent modes that 
satisfy the above boundary conditions. Evidently, we get a zero mode from 
$V$ but not from $\phi$, and so the low-energy theory is pure 4D, $\N=1$
U(1)
theory.

In 6D, we can e.g. compactify on $T^2/Z_3$,  by imposing ($\omega^3$ = 1)
\begin{eqnarray}
V(x_{\mu},z,\bar{z}) &=& V(x_{\mu},\omega z,\bar{\omega} \bar{z}) \\
\nonumber
\phi(x_{\mu},z,\bar{z}) &=& \omega\phi(x_{\mu},\omega z,\bar{\omega} \bar{z})
\end{eqnarray}
together with the periodicity conditions on the torus. 
Again, $\phi = 0$ at the fixed points of the torus, and this projects out the theory to $\N=1$ SYM in 4D. 

The non-Abelian case offers more interesting possibilities, 
since we can combine a gauge transformation with the orbifold symmetry. 
Of course, for chiral theories in even dimensions, we need
to worry about anomaly cancellation in the bulk. A simple anomaly-free
example in say $D=6$  is obtained,
 however, by imagining that we dimensionally reduce from
e.g. seven dimensions where there are no anomalies. The 6D particle 
content is then a vector multiplet and a hypermultiplet in the adjoint 
representation. The 6D pure gauge anomaly is proportional to  
\begin{equation}
tr F^4_{\mbox{\tiny{Adj.}}} - tr F^4_{\mbox{\tiny{Hyper}}}
\end{equation}
which clearly vanishes for a simple hyper in the adjoint rep. (The
gravitational anomalies can be canceled with the Green-Schwarz
mechanism.)

For an amusing example, suppose we start with an $SU(9)$ theory in 6D. 
We will again compactify on $T^2/Z_3$, but this time using the orbifold 
symmetry 
\begin{eqnarray}
V(x_{\mu},z,\bar{z}) = U^{\dagger} V(x_{\mu}, \omega z, \bar{\omega}
\bar{z}) 
U \\
\phi_i(x_{\mu},z,\bar{z}) = \omega U^{\dagger} \phi_i(x_{\mu},\omega z,\bar{\omega} \bar{z}) U
\end{eqnarray}
Where  $U$ is a $9 \times 9$ matrix written in term of $3 \times 3$ blocks:
\begin{eqnarray}
U = 
\begin{pmatrix} 1 & 0 & 0 \\
                0 & \omega & 0 \\
                0 & 0 & \omega^2 \end{pmatrix}
\end{eqnarray}

$V,\phi$ can also be written as a general $9 \times 9$ matrix:
\begin{eqnarray}
V    = \begin{pmatrix}  A & B & C \\
                        D & E & F \\
                        G& H & I \end{pmatrix}, 
 \phi = \begin{pmatrix}  A'& B'& C'\\
                        D' & E' & F' \\
                        G' & H' & I' \end{pmatrix} 
\end{eqnarray}
Now,
\begin{eqnarray}
 U^{\dagger} V U = 
   \begin{pmatrix} A & \omega B  & \omega^2 C \\
                     \omega^2 D & E & \omega F \\
                     \omega G & \omega^2 H & I \end{pmatrix},
\omega U^{\dagger} \phi U =
   \begin{pmatrix} \omega A' & \omega^2 B'  & C' \\
                     D' & \omega E' & \omega^2 F' \\
                     \omega^2 G' & H' & \omega I' \end{pmatrix}
\end{eqnarray}

We see that for the zero modes of $V$, only $(A,E,I)$ survive. This means
that the low-energy theory is $\N=1$ with gauge group $SU(3)^3$. 
On the other hand, from the $\phi_i$, $(C',D',H')$ survive, which transform
under $SU(3)^3$ as $\psi_i \sim (\bar{3},3,1,), (3,1,\bar{3}), (1,\bar{3},3)$. 
There is also a superpotential coupling $\psi_1 \psi_2 \psi_3$ which is 
inherited from the $H^c \phi H$ superpotential term. 
This model 
is just the particle content of ``trinification'', with 3 generations,
and a single large Yukawa coupling.

\section{Localizing chiral fermions} 

It is well-known that it is possible to localize chiral Fermions on 
defects in extra dimensions \cite{dwf}. 
The simplest example are Fermions localized to 
domain walls. Consider a Fermion in 5 dimensions with a spatially varying
mass term $m(x_5)$ (which could for instance arise from a Yukawa coupling 
to a scalar field with a ``kink'' profile in the 5'th dimension). It is
easy to see that if $m(+\infty)>0$ and $m(-\infty) < 0$, then upon solving
for the spectrum of the Dirac operator we find a chiral zero mode with 
wavefunction peaked around the location where the mass term goes through zero. 
Of course, if we attempt to compactify the fifth dimension on a circle, 
then we necessarily have a kink and an anti-kink, and we don't get a chiral 
spectrum in the 4D theory. However, we can combine Fermion localization
with an orbifold to keep the e.g. the left-handed localized zero mode but 
project out the right handed one, as in \cite{Howard}. 

In this section we will supersymmetrize these models and address some
physical questions that arise. Consider 5D theories. Note that 
charged hyper-multiplets in the bulk have a superpotential coupling 
$H^c \phi H$, which is an effective mass term when $\phi$ is non-zero. 
If we can arrange for $\phi$ to vary and change sign from one side of a 
brane to another, then we can localize one of $H,H^c$ to the brane. 
It is easy to arrange for this to happen. 
The simplest example to consider is a 
5D theory with a $U(1)$ gauge field in the bulk, and a 
brane located at $x_5=0$. We will add a Fayet-Iliopoulos term for the gauge
bulk gauge field on the brane. The action is 

\begin{equation}
\mbox{Free 5D action} + \int d^4x \int d^4 \theta2 \zeta V(x,x_5=0)
\end{equation}
The $D$ term is now given by 
\begin{equation}
D = - \partial_5 \Sigma + 2 \zeta \delta(x_5) 
\end{equation}
The most general solution to the $D-$flatness conditions is then 
\begin{equation}
\Sigma (x_5) = \Sigma_0 + \zeta \mbox{sgn}(x_5)
\end{equation}
This is a ``kink'' for $\Sigma$. 
There is moduli space of vacua labeled by $\Sigma_0$. Note that in the 
range $|\Sigma_0| < |\zeta|$, $\Sigma(x_5)$ changes sign as it goes through
the origin, while for $|\Sigma_0| > |\zeta|$, $\Sigma(x_5)$ is
non-vanishing and of the same sign everywhere. 

Now, let us add a bulk Hypermuliplet with charge $+1$ under the $U(1)$. 
Treating the gauge field as a background,the hypermultiplet action becomes
\begin{equation}
\int d x_5 \int d^4 \theta \bar{H} H + \bar{H^c} H^c + \int d^2 \theta 
H^c (\partial_5 - \frac{1}{2} \Sigma(x_5)) H + \mbox{h.c.}
\end{equation}
If we are to have zero modes for $H$ or $H^c$, their wavefunctions 
$\psi(x_5),\psi^c(x_5)$ must satisfy  
\begin{eqnarray}
(\partial_5 - \frac{1}{2} \Sigma(x_5)) \psi &=& 0 \\
(\partial_5 + \frac{1}{2} \Sigma(x_5)) \psi^c &=& 0
\end{eqnarray}
The solutions are trivially 
\begin{eqnarray}
\psi(x_5) = \psi(0) e^{\int_{0}^{x_5} dy \frac{1}{2} \Sigma(y)} = \psi(0) 
e^{\frac{1}{2}(\Sigma_0 + \zeta \mbox{sgn}(x_5)) x_5} \\ 
\psi^c(x_5) = \psi^c(0) e^{-\int_{0}^{x_5} dy \frac{1}{2} \Sigma(y)}
= \psi^c(0) 
e^{-\frac{1}{2}(\Sigma_0 + \zeta \mbox{sgn}(x_5)) x_5} 
\end{eqnarray}
Clearly for $|\Sigma_0| > |\zeta|$, neither of these solutions is
normalizable
and there are no localized chiral zero modes. However, for 
$|\Sigma_0| < |\zeta|$, one (but not the other) of the above two
wavefunctions will be normalizable and we localize a chiral fermion to the 
brane at $x_5 = 0$. 

So, in one region of moduli space $|\Sigma_0| < |\zeta|$, we have a chiral 
zero mode but for $|\Sigma_0| > |\zeta|$ it disappears. How can the net
chirality change as we smoothly move around in moduli space? Mathematically, 
as $|\Sigma_0|\to |\zeta|$, the wavefunction of the chiral zero mode 
spreads out more and more till at $|\Sigma_0| = |\zeta|$ it
it is unnormalizable. Physically, what is going on is also transparent. 
For $|\Sigma_0| < |\zeta|$, there is a normalizable zero mode, and then (
since the Fermions are massive both for $x_5>0$ and $x_5 < 0$) there is
mass gap above which we have the full 5D continuum. Therefore the
low-energy theory is indeed 4-dimensional, and remains that way as we
smoothly
vary $\Sigma_0$. However, as $|\Sigma_0|$ approaches $|\zeta|$, the bulk
Fermion
mass on one side of the brane approaches zero till exactly at $|\Sigma_0| =
\zeta$, the bulk mass term is zero on one side and the low-energy theory 
is not 4D. As we continue to $|\Sigma_0| > |\zeta|$, the bulk Fermion become 
massive in the bulk again. But to an observer on the brane, net chirality has 
been changed. There is of course no contradiction with the usual statement
that net chirality cannot change in 4D, because as we move around in 
moduli space we go through a region where there is no effective 4D
description. This is an elementary analog of the chirality-changing 
transitions in string theory discussed in \cite{EvaShamit}. There, 
chirality changing transitions occured while moving around in moduli space, 
when the effective 4D {\it field theory} description of the 
physics broke down due to the appearance of tensionless strings. 

There are simple variations on the above model. For instance, instead of 
introducing a FI term on the brane, we could introduce a pair of 
chiral fields $X,\bar{X}$ of charge $+1,-1$, 
 which can take arbitrary vevs. Normally, 
this would violate $D-$ flatness, but in this case we simply have 
\begin{equation}
D = -\partial_5 \Sigma + (|X|^2 - |\bar{X}|^2) \delta(x_5)
\end{equation}
and so we have have the same $D-$flat solution for $\Sigma$ as before 
with $\zeta \to (|X|^2 - |\bar{X}|^2)$.

We can also easily discuss compactification and the supersymmetric 
generalization of the models in \cite{Howard}. We will consider an
$S_1/Z_2$ of the model with a $U(1)$ gauge field and a hypermultiplet in
the bulk. The orbifold symmetry is 
\begin{eqnarray}
V(x,x_5)  & \to & V(x,-x_5) \\
\phi(x,x_5) & \to  & - \phi(x,-x_5) \\
H(x,x_5) &\to& H(x,-x_5) \\
H^c(x,-x_5) &\to& -H^c(x,-x_5)
\end{eqnarray}
Furthermore, on the orbifold fixed point at $x_5=0$ we will write down a 
FI term, while on the fixed point at $x_5 = L$ we will put a pair of 
chiral fields $Y,\bar{Y}$ of charge $+1,-1$. The $D-$ flatness 
condition is now 
\begin{equation}
\partial_5 \Sigma + 2 \zeta \delta(x_5) + (|Y|^2 - |\bar{Y}|^2)
\delta(x_5 - L) = 0
\end{equation}
The general solution to this equation is 
\begin{equation}
\Sigma(x_5) = \Sigma_0 + \zeta \mbox{sgn}(x_5) + \frac{1}{2}(|\chi|^2 - 
|\bar{\chi}|^2) \mbox{sgn}(x_5 - L).
\end{equation}
However, in order to be able to find a solution invariant under the 
orbifold symmetry we must have 
\begin{equation}
\Sigma_0 = 0, |\chi|^2 - |\bar{\chi}|^2 + 2 \zeta = 0.
\end{equation}
So the $\Sigma$ modulus has been projected out, and the second condition
is just the usual $D$-flatness condition in 4D, (as it had to be from 
the low-energy point of view). 

Now, we can look at what happens to the hypermultiplets in this
background. The solutions for the zero mode wavefunctions of $H,H^c$ 
are 
\begin{equation}
\psi = A e^{\frac{\zeta}{2} x_5},\psi^c = B e^{-\frac{\zeta}{2} x_5}
\end{equation}
however, by the orbifold symmetry, the zero mode of $H^c$ must vanish at
the orbifold fixed points, so $B$ must vanish and there is therefore no 
zero mode for $H^c$. The zero mode for $H$, on the other hand, can be 
localized at either $x_5=0$ or $x_5=L$ depending on the sign of $\zeta$.

Finally, we can replace the FI term on the fixed point at $x_5=0$ by
another
pair of chiral multiplets $X,\bar{X}$ of charge $\pm 1$. Then everything
goes through the same with $\zeta \to (|X|^2 - |\bar{X}|^2)$. There is 
a moduli space of solutions corresponding to the usual $D-$flat space in
4D. As we move along this moduli space, the chiral fermion wavefunction 
can shift from 
being localized around $x_5=0$, to having a flat wavefunction, to being
localized around $x_5=L$.

\section{Anomalies and super-Chern-Simons Theory}
We have shown how to localize chiral fermions in a fifth direction. It is
then natural to ask what happens with anomalies in such a theory. 
In the case of domain-wall fermions, it is well-known that the apparent
anomaly due to the localized chiral zero mode is canceled by the variation
of a Chern-Simons term in the bulk \cite{Anomaly}. We will not repeat the whole story
here. The important point is that the variation of the Chern-Simons action
on a manifold $M$,
under a gauge transformation $\delta A = d \Lambda$, is 
\begin{equation}
\delta \int_M A \wedge F \wedge F = \int_M d \Lambda \wedge F \wedge F 
= \int_M d(\Lambda  F \wedge F) = \int_{\partial M} \Lambda F \wedge F
\end{equation}
and the integral over the boundary has precisely the form of a 4D anomaly. 
As such, it can cacel the 4D anomaly induced by
fields living at the boundaries. 

What we would like to do here is show how to supersymmetrize the
5D Chern-Simons term in our 4D superspace formalism. Note that the usual
 Chern-Simons term contains $A_5 F \tilde{F}$. It is easy to see that this
term must come from the term $\int d^2 \theta \phi W W$. However, note that
the gauge coupling of the 5D YM theory can be absorbed by shifting
$\phi$. This leads us to guess that 5D SYM + 5D super-Chern-Simons theory
is actually on the moduli space of pure 5D super-Chern-Simons. As we will
see, this is indeed correct. We will therefore only construct the action
for pure super-Chern-Simons theory. 

The piece of the action we have so far, $\int d^2 \theta \phi W W$, is
clearly not fully gauge invariant, nor fully 5D Lorentz-invariant. 
It is not difficult to find the correct combination of terms required. The
correct action for super-Chern-Simons theory, on an interval between
$x_5=[y_1,y_2]$, is 
\begin{eqnarray}
&&S^{\mbox{\tiny{5D CS}}} = \int d^4 x \int_{y_1}^{y_2} dx_5 \int d^2 \theta \phi W W + \text{h.c} \nonumber \\
&& - \frac{\sqrt{2}}{3} \int d^4 \theta (\partial_5 V D_\alpha V W^\alpha -
V D_\alpha \partial_5 V W^\alpha) + \text{h.c} \nonumber \\ 
&& - \frac{1}{3} \int d^4 \theta (\sqrt{2} \partial_5 V - (\phi +
\bar{\phi}))^3
\end{eqnarray}
The bosonic part of the component action is:
\begin{eqnarray}
S^{\mbox{\tiny{5D CS}}}_{\text{bosonic}} = \int d^4 x \int_{y_1}^{y_2} dx_5 -\frac{1}{2\sqrt{2}} \epsilon^{MNOPQ} A_M F_{NO}F_{PQ} + \frac{1}{\sqrt{2}} \Sigma F^{MN}F_{MN} + \sqrt{2} \Sigma \partial_M \Sigma \partial^M \Sigma
\end{eqnarray}

It is easy to verify with this action that under the gauge transformation  
$\delta \phi = \partial_5 \Lambda, \delta V = \Lambda + \bar{\Lambda}$, the 
above action has a variation
\begin{equation}
\delta S^{5DCS} = \int d^4 x \int d^2 \theta (\Lambda W W)(y_2) - (\Lambda W W)(y_1) + \text{h.c}
\end{equation}
which is the full supermultiplet of chiral anomalies on the boundaries at
$y_1,y_2$.

\section{Conclusions}
In this paper we have given the rules for constructing globally
supersymmetric Lagrangians from $\D=5$ to $\D=10$ dimensions in the familiar 
$\N=1,\D=4$ superspace. This makes it easy to do explicit supersymmetric
model-building in extra dimensions, in particular allowing us to couple
bulk fields to fields localized on 3-branes with ease. We illustrated
the utility of the formalism with a number of simple examples. It would be
interesting to explore some generalizations of these examples in
detail. For instance, it would be nice to generalize the supersymmetric
localization of chiral fermions to higher dimensions. 

There are also a number of possible extensions of the ideas in this paper
that we have not touched on. For instance, when we have dynamical
branes which fluctuate, with finite tension, there are massless scalar fields 
living on the branes which are the goldstone bosons of spontaneously broken
translational invariance. They non-linearly realize the full translational
symmetry of the theory \cite{Raman}. In the case where the brane preserves some SUSY,
these goldstone modes must fall into supermultiplets, and it would be
interesting to know how to systematically construct Lagrangians
non-linearly realizing the full SUSY. A related possible 
application of our formalism is the
construction of BPS ($\N=1$ preserving) solitons in the higher-dimensional 
theory. These would be $F-$ and $D-$ flat solutions with non-trivial
variation of fields in the extra dimensions. 
Finally, it would of
be desirable to extend our formalism to the case of supergravity.
While a full treatment may be difficult, the case of
linearized supergravity could be tractable, and would already contain much
of the interesting physics.

\section{Acknowledgements}
We would like to thank L. Hall, M. Luty, Y. Nomura, U. Varadarajan and N. Weiner for useful
discussions. We also thank A. Sagnotti, W. Siegel, S.Gates and N. Berkovitz
for informing us of the early reference \cite{siegel}.
This work was supported in part by the D.O.E. under Contracts 
DE-AC03-76SF00098, and in part by the National Science Foundation under
grant PHY-95-14797. The work of N.A-H is also supported by the Alfred P. Sloan
foundation and by the David and Lucile Packard Foundation. T. Gregoire is
also supported by an NSERC fellowship.

\end{document}